\documentclass[10pt,letterpaper,twocolumn]{article} 

\newcommand{\be}{\begin{equation}}
\newcommand{\ee}{\end{equation}}
\newcommand{\bea}{\begin{eqnarray}}
\newcommand{\eea}{\end{eqnarray}}
\newcommand{\beaa}{\begin{eqnarray*}}
\newcommand{\eeaa}{\end{eqnarray*}}
\newcommand{\ben}{\begin{enumerate}}
\newcommand{\een}{\end{enumerate}}
\newcommand{\bi}{\begin{itemize}}
\newcommand{\ei}{\end{itemize}}

\newcommand{\df}{{\rm d}}
\usepackage{BLANK}
\usepackage[draft]{hyperref}
\usepackage{amsmath}
\usepackage{amssymb}
\begin{document}

\twocolumn[ 

\title{Nonlinear propagation in multi-mode fibers in the strong coupling regime}

\author{Antonio Mecozzi,$^1$ Cristian Antonelli,$^1$ and Mark Shtaif$^2$	}
\address{$^1$Dipartimento di Ingegneria Elettrica e dell'Informazione, Universit\`a dell'Aquila, 67100 L'Aquila, Italy \\
$^2$School of Electrical Engineering, Tel Aviv University, Tel Aviv, Israel 69978}

\begin{abstract}
In spite of the massive interest that the generalized Manakov equation has attracted in the past two decades, no physical system which is quantitatively described by this equation has been reported so far. In this paper we show that propagation in a group of degenerate modes of a multi-mode optical fiber satisfies this equation in the presence of random mode coupling. Consequently, this is the first reported physical system that admits true  multi-component soliton solutions. The reported formalism constitutes the starting point for future studies of nonlinear effects in multi-mode fiber transmission.
\end{abstract}

] 

\noindent Several years after the publication of the famous Zakharov and Shabat paper \cite{Zakharov} which showed that the scalar nonlinear Schr\"{o}dinger equation (NLSE) is integrable, Manakov \cite{Manakov} discovered that a special form of a two-component vector NLSE in which the nonlinearity is isotropic shares the same property. However, no practical physical system was described by this equation until Wai, Menyuk and Chen \cite{Wai0} found that Manakov's equation accurately describes the propagation of a polarized optical field in single-mode birefringent optical fibers in the presence of random mode coupling. This implied that such fibers support the existence of vector solitons, as predicted by Manakov almost two decades earlier.

Recently, with fiber-communications exhausting the capacity of single-mode fibers, multi-mode fibers are being considered for transmission, with the purpose of increasing the overall fiber-communications throughput through spatial multiplexing. This new paradigm calls for the extension of the existing theoretical framework to the multi-mode fiber case. Random coupling of degenerate or quasi degenerate modes is a distinctive feature of these fibers \cite{Ryf,Salsi}. We have addressed the linear problem of mode coupling and modal dispersion in multi-mode fibers in a recent paper \cite{Antonelli}. In this paper we focus on nonlinear propagation regime, where we show that the evolution of the electric field in a group of degenerate modes of a multi-mode optical fiber is described by a generalized multi-component Manakov equation \cite{Makhankov}. The central feature of the generalized Manakov equation is that it is integrable \cite{Makhankov} and hence supports the propagation of solitons. Those are particle-like waveforms that remain unchanged in the process of propagation and preserve their identity when colliding with each other. It is the latter property that distinguishes between solitons and generic solitary waves, whose shape is also unaltered by propagation, but becomes corrupted when two, or more, such waveforms collide. Interestingly, solitary waves in certain multi-mode optical systems were shown to be possible using specific combinations of  physical parameters \cite{Crosignani,Musslimani,Krolikowski,Assanto,Silberberg}. Yet, unlike the case which we present here, these systems are modeled by non-integrable equations and hence they do not support true soliton propagation.

We consider propagation over a group of degenerate modes in a multi-mode optical fiber. The degenaracy of the modes implies that fiber imperfections in the form of mechanical stress or manufacturing distortions produce strong random coupling between them on a length-scale typically much shorter than the length-scale of the nonlinear evolution \cite{RyfECOC}. On the other hand, the distinctive difference in wave vector between non degenerate groups of modes makes coupling between them significantly smaller and hence we neglect it in our analysis. The four degenerate LP$_{11}$ modes \cite{Ryf} of a step-index fiber are a relevant example for a group of degenerate modes of the kind that we consider here. Larger groups of degenerate modes can be encountered in situations where multiple multi-mode fibers are combined in a multi-core fiber structure \cite{Randel}.

Starting from the coupled NLSE and assuming random mode coupling, we generalize the standard Manakov equation \cite{Wai0} to the multi-mode case. We verify the accuracy of the generalized Manakov equation and demonstrate the existence of multidimansional vector solitons, by solving the complete set of coupled NLSE numerically.

The electric field in a group of $N$ degenerate spatial modes is represented by a $2N$-dimensional complex valued vector $\vec E(z,t)$, which is constructed by stacking the Jones vectors of the $N$ individual spatial modes one on top of the other. The term spatial mode is used here and throughout the paper to refer to the set of two polarization modes sharing, in the weakly guiding fiber approximation \cite{Gloge}, the same lateral field profile. The components of $\vec E(z,t)$ then satisfy the following set of coupled NLSE \cite{Agrawal,Poletti}
\bea && \frac{\partial E_j}{\partial z} = i \sum_{m} \beta_{jm}^{(1)} E_m - \sum_m \beta_{jm}^{(1)} \frac{\partial E_m}{\partial t} \nonumber \\ && - i  \sum_m \frac{\beta_{jm}^{(2)}} 2 \frac{\partial^2 E_m}{\partial t^2} + i \gamma \sum_{hkm} C_{jhkm} E_h^* E_k E_m, \label{10} \eea
which, in a vector form, can be re-expressed as
\bea \frac{\partial \vec E}{\partial z} &=& i \mathbf B^{(0)} \vec E  - \mathbf B^{(1)} \frac{\partial \vec E}{\partial t} - i  \frac{\mathbf B^{(2)}} 2 \frac{\partial^2 \vec E}{\partial t^2}\nonumber\\&& + i \gamma \sum_{jhkm} C_{jhkm} E_h^* E_k E_m \hat e_j, \label{21} \eea
where $\mathbf B^{(i)}(z)$, $i=0,1,2$ are $2N\times2N$ Hermitian matrices and where we use $\hat e_j$ with $j=1,\dots, 2N$ to denote the set of complex orthogonal unit vectors used to represent the electric field.

The fourth term on the right-hand side of Eqs. (\ref{10}) and (\ref{21}) represents the coupling induced by the Kerr nonlinearity of the fiber. The  coefficient $\gamma = \omega_0 n_2 / c A_\mathrm{eff}$  is identical to the usual nonlinearity coefficient appearing in the scalar NLSE \cite{Agrawal} of a single mode fiber, where $n_2$ is the Kerr coefficient of glass, $c$ is the speed of light in vacuum, and $A_\mathrm{eff}$ is the effective area of the fundamental mode at central frequency $\omega_0$. The dimensionless constants $C_{jhkm}$ depend on the details of the spatial mode profiles and are obtained as follows \cite{Poletti}. Defining
\bea  D_{jhkm}^{(1)} &=& \frac{ \int\df x\df y (\vec{F}_j^* \cdot \vec{F}_m )(\vec{F}_h^* \cdot \vec{F}_k ) }{N_j N_h N_k N_m},  \\
D_{jhkm}^{(2)} &=& \frac{ \int\df x\df y (\vec{F}_j^* \cdot \vec{F}_h^* )(\vec{F}_m \cdot \vec{F}_k ) }{N_j N_h N_k N_m}, \\
N_n^2 &=& \int \df x \df y |\vec{F}_n|^2 n_f, \eea
where $\vec{F}_n(x,y)$ describes the lateral mode profiles, and $n_f(x,y)$ is the refractive index, we obtain
\bea
C_{jhkm} = A_0 [2 \, D_{jhkm}^{(1)}+D_{jhkm}^{(2)}]/3,\label{24.5}
\eea
with $A_0^{-1} = [2 D_{1111}^{(1)}+D_{1111}^{(2)}]/3 = D_{1111}^{(1)}$, so that $C_{1111} = C_{2222} = 1$, where the indices 1 and 2 are used to denote the two polarizations of the fundamental mode. Note that $A_0 \simeq n_{\mathrm{eff}}^2 A_{\mathrm{eff}}$, where $n_{\mathrm{eff}}$ is the effective refractive index of the fundamental mode at the central frequency $\omega_0$.

The first three terms on the right-hand side of Eqs. (\ref{10}) and (\ref{21}) account for linear propagation and coupling \cite{Antonelli}. 
While in the ideal case the matrices $\mathbf B^{(i)}(z)$ ($i=0,1,2$) are proportional to the identity, unavoidable position dependent perturbations result in the presence of off-diagonal terms, producing linear coupling between the various modes. The strongest coupling results from the term proportional to $\mathbf B^{(0)}$, and in most cases of practical interest, its characteristic length-scale is shorter by orders of magnitude than the length-scale that characterizes the nonlinear evolution \cite{Agrawal}. Under these conditions the orientation of the electric field vector, defined by $\vec E/|\vec E|$, uniformly samples the space of possible orientations within the length-scale of the nonlinear evolution. Hence, field propagation can be well approximated by averaging the nonlinear terms with respect to the electric field's orientation. As can be anticipated based on symmetry and dimensionality arguments, the averaged nonlinear term of (\ref{21}) must reduce to the form
\bea \sum_{jhkm} C_{jhkm} E_h^* E_k E_m \hat e_j=\kappa|\vec E|^2\vec E.\label{30.5}\eea
where $\kappa$ is a dimensionless parameter that depends only on the nonlinear coupling coefficients $C_{jhkm}$. Following the steps described in the appendix, we show that
\bea \kappa=\sum_{jhkm}C_{jhkm}\frac{\delta_{hk}\delta_{jm}+\delta_{hm}\delta_{jk}}{2N(2N+1)}, \label{41}\eea
where $\delta_{ij}$ is Kronecker's delta function. Equation (\ref{21}) can thus be reduced to
\be \frac{\partial \vec E}{\partial z} = i \mathbf B^{(0)} \vec E  - \mathbf B^{(1)} \frac{\partial \vec E}{\partial t} - i  \frac{\mathbf B^{(2)}} 2 \frac{\partial^2 \vec E}{\partial t^2} + i \gamma \kappa |\vec E|^2\vec E. \label{91} \ee
This equation is a $2 N$ component generalization of the well-known Manakov-PMD equation describing the propagation in a single-mode optical fiber with random polarization coupling \cite{Wai}. Indeed, when $N=1$, we have $C_{1221} = C_{2112} = 2/3$ and $C_{1111} = C_{2222} = 1$ \cite{Agrawal}, yielding $\kappa = 8/9$, as expected \cite{Wai}.
The non-diagonal terms in the matrices $\mathbf B^{(1)}$ and $\mathbf B^{(2)}$ are due to the frequency dependence of the mode coupling, which is typically only a small correction to the coupling contained in the frequency independent part $\mathbf B^{(0)}$. By neglecting these terms and  transforming Eq. (\ref{91}) to a reference frame evolving with $\mathbf B^{(0)}$, the final form of the generalized Manakov equation is obtained
\be \frac{\partial \vec E}{\partial z} = - \beta' \frac{\partial \vec E}{\partial t} - i  \frac{\beta''} 2 \frac{\partial^2 \vec E}{\partial t^2} + i \gamma \kappa |\vec E|^2\vec E. \label{92} \ee
The terms $\beta'$ and $\beta''$ are the inverse group velocity and the dispersion coefficient, respectively. The form of Eq. (\ref{92}) is identical to that of the familiar Manakov equation \cite{Manakov}, except that it describes the propagation of signals through $2N$ degenerate propagation modes of a multi-mode fiber.


The generalized Manakov equation (\ref{92}) is integrable by the inverse scattering transform \cite{Makhankov}, and hence 
it admits the propagation of solitons. Unlike generic solitary waveforms that were shown to exist in multi-mode fibers with certain parameter combinations \cite{Crosignani}, true solitons can only exist in the strong coupling regime, where propagation is accurately described by Eq. (\ref{92}). Thus, in order to test the accuracy of Eq. (\ref{92}) in describing multi-mode propagation, we verify that the fundamental soliton waveform, which is a rigorous analytical solutions of Eq. (\ref{92}), indeed forms a solitary solution when the coupled NLSE (\ref{10}) are solved numerically in the strong mode-coupling regime. In addition, we check that when two such waveforms are launched into different fiber modes and at different central optical frequencies, they collide elastically without leaving a trace in the form of a dispersive wave. As noted earlier, this phenomenon is a distinctive feature of solitons \cite{Zakharov} distinguishing them from generic solitary waves. We find that Manakov solitons can be observed when the length-scale characterizing the correlation of mode coupling is smaller than the soliton length $L_s = \tau^2/|\beta''|$ (related to the soliton period $z_s$ by $z_s = \pi L_s/2$).

In the simulations we consider propagation in the two degenerate LP$_{11}$ modes of a step-index optical fiber, LP$^{(a)}_{11}$ and LP$^{(b)}_{11}$ \cite{Sillard}. We used a core radius of 7.5$\mu$m, a core refractive index of 1.4621, a refractive index step of $9.7 \times 10^{-3}$ and a dispersion coefficient $\beta'' = - 25$ps$^2$/km. The nonlinear coefficient in our computations was $n_2 = 2.6 \times 10^{-20}$m$^2$W$^{-1}$, corresponding to $\gamma \simeq 0.835 $ W$^{-1}$km$^{-1}$, $\kappa \simeq 0.76$, and the effective area was of $126\mu$m$^2$ for the fundamental mode.

\begin{figure}[t!]
\centering\includegraphics[width=0.8\columnwidth]{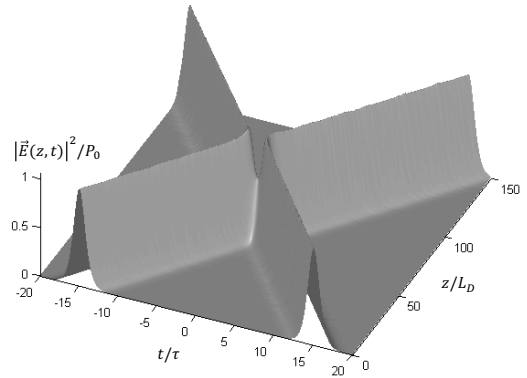}
\caption{Collision of two solitons initially in two orthogonal modes in the case of full coupling, and initial angular frequency separation $\Delta \omega = 0.4 / \tau$. The time axis is normalized to the soliton time $\tau$, the spatial axis to the soliton length $L_s$ and the vertical axis to the soliton peak power $P_c$. The ratio of the correlation length to the soliton length is $L_c/L_s \simeq 2.8 \times 10^{-3}$.}\label{fig1}
\end{figure}

In Fig. 1 we consider the case in which we launch the waveform describing the fundamental soliton of the Manakov equation (\ref{92}), $P_0^{1/2} \, \mathrm{sech}(t /\tau)$ with $P_0=|\beta''|/(\gamma\kappa\tau^2)$, into each of the two modes LP$^{(a)}_{11}$ and LP$^{(b)}_{11}$.
The two launched soliton waveforms are separated by $T=30\tau$ in time and by $\Delta\omega=0.4/\tau$ in frequency. As random coupling of polarizations within each spatial mode is always very fast \cite{Wai0}, the launch polarization states of the two waveforms is immaterial. We assume a soliton half-width $\tau_h = 17$ps corresponding to $\tau = 30$ps. The soliton length is $L_s \simeq 36$km and the peak power $P_0 = 44$mW. The correlation length of the mode coupling was taken to be $L_c = 100$m ($L_c/L_s \simeq 2.8 \times 10^{-3}$). It is evident that the launched waveforms are indeed solitary solutions in this regime and the collision between them is elastic, as expected from true soliton behavior. The collision dynamics shown in Fig. \ref{fig1} is indistinguishable from that obtained integrating the Manakov equation with the same initial conditions.

In Fig. 2 we consider the regime in which the spatial modes are perfectly isolated so that no coupling between them occurs, although full coupling occurs between polarizations in each mode. In this case we launch the same sech-waveform, but with  $P_0=|\beta''|/(8\gamma C_{3333}\gamma\tau^2/9)$, where $C_{3333}\simeq 1.07$ corresponds to self-phase modulation in the LP$_{11}$ modes (see definitions around Eq. (\ref{24.5}), where 3 is the index of mode LP$^{(a)}_{11}$). Since the launched waveform is solitary for the individual modes LP$^{(a)}_{11}$ and LP$^{(b)}_{11}$, the pulses propagate unperturbed until their collision. Yet, upon collision they stick to each other, forming a new (and non-solitary) pulse, while some energy is lost in the form of dispersive radiation. This behavior is in clear contrast with the elastic collision observed in Fig. 1, which corresponds to the strong coupling regime, in which the generalized Manakov equation holds.

In Figs. 3 and 4 we consider cases where the correlation length of the random mode coupling is $L_c=10$km ($L_c/L_s \simeq 2.8 \times 10^{-1}$) and $L_c=100$km ($L_c/L_s \simeq 2.8$), respectively.
In the case of $L_c=10$km shown in Fig. 3, the pulse evolution is similar to that observed in the strong coupling case of Fig. 1, although some perturbations to the propagating waveforms can be observed due to the insufficient averaging of the random mode coupling in this regime. These perturbations are further exacerbated when the correlation length of the mode coupling is extended to $100$km, as shown in Fig. 4.
In both Figs. 3 and 4, the launched waveforms were Manakov-equation solitons, identical to those used in Fig. 1.

\begin{figure}[t!]
\centering\includegraphics[width=0.8\columnwidth]{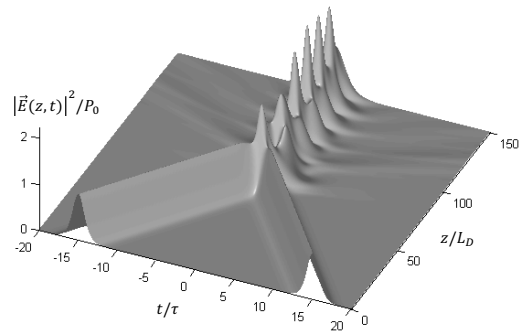}
\caption{Collision of two solitons initially in two orthogonal modes in the case of absence of coupling, and initial angular frequency separation $\Delta \omega = 0.4 / \tau$. The time axis is normalized to the soliton time $\tau$, the spatial axis to the soliton length $L_s$ and the vertical axis to the soliton peak power $P_u$}\label{fig2}
\end{figure}
\begin{figure}[t!]
\centering\includegraphics[width=0.8\columnwidth]{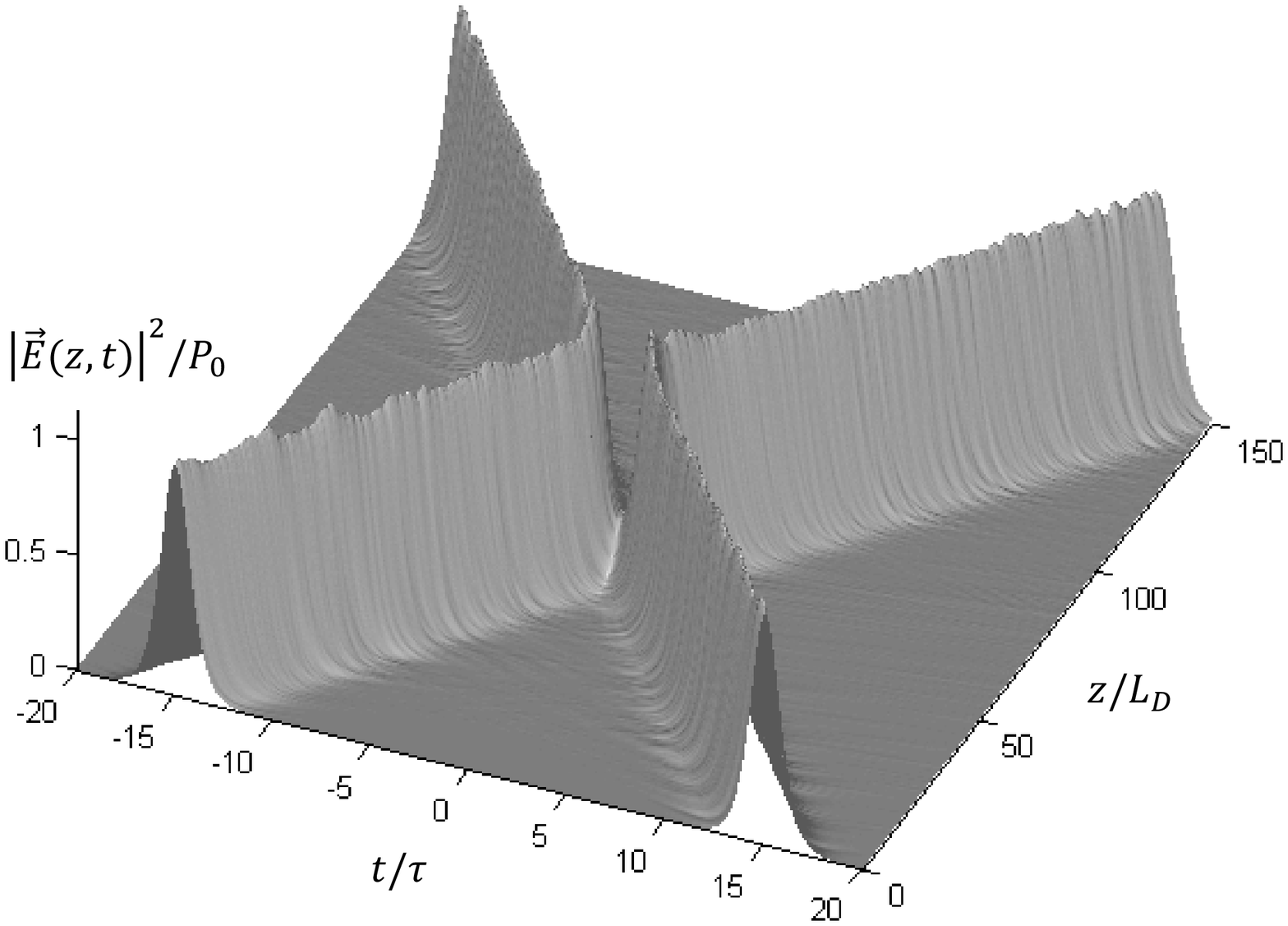}
\caption{Collision of two solitons initially in two orthogonal modes in the same condition of Fig. \ref{fig1}, with ratio of the correlation length to the soliton length of $L_c/L_s \simeq 0.28 \times 10^{-1}$.}\label{fig3}
\end{figure}
\begin{figure}[t!]
\centering\includegraphics[width=0.8\columnwidth]{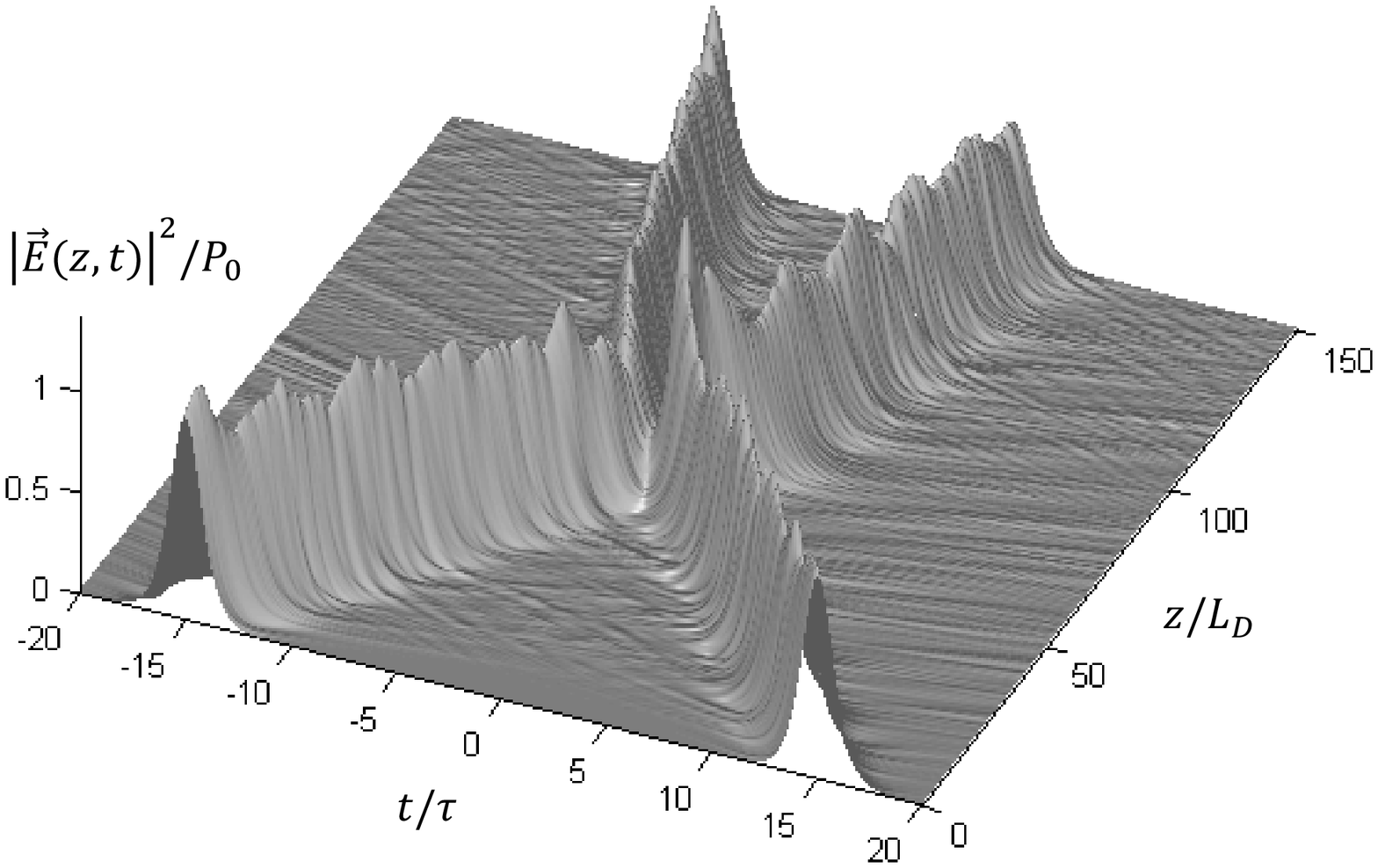}
\caption{Collision of two solitons initially in two orthogonal modes in the same condition of Fig. \ref{fig1}, with ratio of the correlation length to the soliton length of $L_c/L_s \simeq 2.8$.}\label{fig4}
\end{figure}

The generalized Manakov equation (\ref{92}) constitutes a powerful tool for understanding the nonlinear dynamics in multi-mode optical fibers. As in the case of its two-component predecessor, describing polarization dynamics in single-mode fibers \cite{Wai0}, it constitutes a starting point for most analytical descriptions of nonlinear propagation phenomena \cite{Mollenauer}. In addition, as it incorporates the effect of random coupling analytically, it can be used for an effective simplification of numerical studies of signal propagation \cite{Marcuse}. Possible applications include the analysis of inter and intra-channel interference in fiber-communications systems using multiplexing in both wavelength and space. Recently, such systems, deployed over multi-mode \cite{MultiMode} and multi-core \cite{Multicore} optical fibers are being hectically studied with the hope of providing an economically viable path for overcoming the capacity crunch experienced in the field of fiber-communications \cite{Chraplyvy}.

In order to show that in the presence of random mode coupling the coupled NLSE (\ref{10}) are equivalent to the generalized Manakov equation (\ref{92}) we solved (\ref{10}) numerically and showed that it supported the fundamental soliton waveforms, which are exact analytical solutions of (\ref{92}). To demonstrate that those were indeed solitons and not generic solitary waves, we verified that collisions between pairs of such waveforms were elastic, namely such that maintain the identity of the individual solitons and produce no dispersive waves. The accuracy of the Manakov equation in describing the field evolution was shown to be excellent when the correlation length of the mode coupling was smaller than the nonlinear length by an order of magnitude, or more.

The fact that light propagating through degenerate modes in a multi-mode fiber satisfies the generalized Manakov equation can be exploited for the prediction of multiple general properties of signal propagation. An important example is that the isotropy of the Manakov equation implies that collisions between two polarized pulses (i.e. pulses whose orientation $\vec E(t)/|\vec E(t)|$ is time independent) can always be considered as occurring in a two dimensional (complex) subspace of the $2N$ dimensional space spanned by the entire set of modes \cite{Tsuchida, Soljacic}. Consequently, the description of two-pulse interactions becomes identical to the description of two-pulse interactions in single-mode fibers, allowing use of the conventional Stokes-space representation \cite{Gordon&Kogelnik}. Then, it can be shown that pulse collisions correspond to the Stokes vector's representing each of the pulse precessing about each other \cite{Mollenauer,Mecozzi_OE}. When the colliding pulses are orthogonal to each other (implying that their corresponding Stokes vectors are antiparallel to one another) their orientations in the $2N$-dimensional space of the electric field remain unaltered. Conversely, when the input pulses are not orthogonal, their Stokes vectors are not aligned with each other and precession in Stokes space implies that the pulses' orientations are modified by the collision, so that the distribution of the overall optical power among the various fiber modes changes. This effect, which could not be easily predicted using Eq. (\ref{10}), may be of importance to space and wavelength multiplexed systems when optical nonlinearities are non negligible.

We showed that propagation in a degenerate group of randomly coupled spatial modes of a multi-mode optical fiber is the first reported physical scenario which is described by the generalization \cite{Makhankov} of the famous Manakov equation \cite{Manakov}, and hence admits true vector soliton solutions. The key parameter of the equation describing the nonlinear effect was expressed in terms of the standard parameters of a generic optical fiber. This equation constitutes a starting point for analytical and numerical studies of nonlinear effects in multi-mode transmission.
Its importance is emphasized by the massive recent interest in spatially multiplexed transmission using multi-mode and multi-core optical fibers.

\section*{Appendix}
In order to derive Eq. (\ref{41}), we perform
scalar multiplication of both sides of Eq. (\ref{30.5}) by $\vec E$ and average with respect to the field's orientation. This yields
%
\be \kappa=\sum_{jhkm} C_{jhkm} Q_{jhkm},\ee
where $\mathbb{E}$ denotes statistical averaging and $Q_{jhkm}=\mathbb{E} \left[ E_h^* E_k E_m E_j^*\right]/|\vec E|^{4}$. Since $\vec E$ can be considered as a constant modulus vector whose orientation is uniformly distributed, the statistical average can be performed as follows. We introduce an auxiliary random vector $\vec X$ with $2N$ complex, statistically independent components. The real and imaginary parts of each component $X_i$ of $\vec X$ are statistically independent standard Gaussian variables having zero-mean and unit variance. It can be argued that $Q_{jhkm}=\mathbb{E}\left[ X_h^* X_k X_m X_j^*\left||\vec X|^2\right.\right]/|\vec X|^{4}$. Multiplying both sides by $|\vec X|^4$ and performing another average (with respect to the square modulus $|\vec X|^2$), we find that
$$Q_{jhkm}=\frac{\mathbb{E}\left[ X_h^* X_k X_m X_j^*\right]}{\mathbb{E}\left[|\vec X|^{4}\right]}.$$
Since all components of $\vec X$ are statistically independent standard complex Gaussian variables, the numerator is $4\left(\delta_{hk}\delta_{jm}+\delta_{hm}\delta_{jk}\right)$. The denominator is the second moment of a chi-square random variable having $4N$ degrees of freedom and hence its value is $4N(4N+2)$. Using these results, the final form of Eq. (\ref{41}) readily follows.

\end{document}